\documentstyle[preprint,aps,epsf,floats]{revtex}

\tighten

\def\OMIT#1{{}}

\def\cb{{\cal B}}
\def\cbb{{\overline{\cal B}}}

\def\hlittle{{h}}

\begin{document}

\preprint{\vbox{
\hbox{NT@UW-02-006}
}}

\title{Hadronic Parity-Violation on the Lattice}
\author{{\bf Silas R. Beane}  and {\bf Martin J. Savage}}
\address{Department of Physics, University of Washington, \\
Seattle, WA 98195. }

\maketitle

\begin{abstract} 
  
  We motivate lattice QCD studies of the parity-violating pion-nucleon 
coupling constant and extend flavor-conserving hadronic parity-violation 
from QCD to partially-quenched QCD. The
  parity-violating pion-nucleon coupling and the anapole form factor (and
  moment) of the proton are computed to one-loop order in the 
partially-quenched chiral
  expansion.
For the parity-violating pion-nucleon interaction, 
we include the contributions from total derivative operators
necessary to match the kinematics that will be used in lattice simulations.
\end{abstract}

\vfill\eject

\section{Introduction}

While flavor-changing parity-violating (PV) interactions are well 
understood theoretically and a
great deal of precise data exists, knowledge of flavor-conserving
parity-violation is rather primitive. Flavor-conserving
parity-violation continues to be an area of intense investigation
in the nuclear physics community. 
Its study is presently serving both to uncover the structure of the
nucleon in electron-scattering experiments such as SAMPLE\cite{sample}, and to
determine PV flavor-conserving couplings between pions
and nucleons\cite{cesium,Fluorine}. 
The
problems that are encountered in this sector are both
experimental and theoretical.  On the experimental side, the 
PV signals, unlike those in flavor-changing processes, appear as small
deviations in either a strong or an electromagnetic process, such as PV in
$ep\rightarrow ep$, or in the circular polarization of the $\gamma$-ray emitted
in $^{18}F^* \rightarrow ^{18}F \gamma$~\cite{Fluorine}.  The current situation
is somewhat confused by the fact that measurements of parity-violation in 
atoms and nuclei do not give rise to a consistent set of couplings
between hadrons~\cite{Haxton:2001ay}. However, it is important to keep in 
mind that many of the
``experimental'' determinations of these couplings require theoretical inputs
with varying degrees of reliability.  Recently, it has been reemphasized that
measurements of PV observables in the single-nucleon sector would significantly
ameliorate the situation by eliminating many-body
uncertainties~\cite{BSpv,Chpv}.  Despite the inherent difficulty of such
experiments, there are ongoing efforts to measure PV processes in systems with
only one or two nucleons, such as the angular-asymmetry in 
$\vec np\rightarrow d\gamma$~\cite{snow,Hyun,Savage:2000iv}. 
Such measurements should provide a reliable
determination of the leading-order (LO), momentum-independent weak $\pi NN$ 
coupling constant,
$h_{\pi NN}^{(1)}$.

On the theoretical side, despite heroic efforts to
model~\cite{DDH,Henley} hadronic matrix elements of the four-quark
operators that appear in the low-energy effective theory of the standard model,
there are no reliable calculations of the PV couplings between hadrons. A first
principles calculation of $h_{\pi NN}^{(1)}$ in lattice QCD would therefore be
extremely welcome.  This would require a lattice QCD simulation of a correlator
with three hadronic sources interacting via a four-quark operator.
Unfortunately, chiral symmetry does not allow one to relate the $\pi NN$
correlator to a correlator without the pion. On the bright side, the structure
of the four-quark weak Hamiltonian requires a flavor change in the nucleon and
therefore there are no disconnected diagrams to be computed on the lattice.

A standard difficulty in reliably computing $h_{\pi NN}^{(1)}$
on the lattice is that present computational limitations necessitate the use of
lattice quark masses that are significantly larger than those of nature. In
order to make a connection between lattice calculations in the foreseeable
future and nature, an extrapolation in the quark masses is required.  Chiral
perturbation theory ($\chi$PT) provides a systematic description of low-energy
QCD near the chiral limit, and this technique has been extended to describe
both quenched QCD (QQCD)~\cite{Sharpe90,S92,BG92,LS96,S01a} and
partially-quenched QCD (PQQCD)~\cite{Pqqcd,SS01}. The hope is that future
lattice simulations can be performed with quark masses that are small enough to
guarantee a convergent chiral expansion, thus allowing a meaningful
extrapolation to the quark masses of nature. Recently, it has been shown how to
include the low-lying octet and decuplet of baryons~\cite{CS01a,BS02b} into
partially-quenched chiral perturbation theory
(PQ$\chi$PT)~\cite{Pqqcd,SS01}.  In this work we will present 
$h_{\pi NN}^{(1)}$ and the anapole moment of the proton at one-loop level in
$SU(4|2)_L\otimes SU(4|2)_R$ PQ$\chi$PT with two non-degenerate light quarks.
Given that lattice simulations have on-shell particles in both the initial
and final states,  extraction of $h_{\pi NN}^{(1)}$  requires the
injection of energy at the weak vertex.  This energy non-conservation
modifies the effective field theory description of the three-point function 
by requiring the inclusion of operators that are total derivatives.
We present a calculation of $h_{\pi NN}^{(1)}$ at the one-loop level 
both for the case where energy is conserved and the pion is off its mass-shell
with $P_\pi^\mu=0$, and 
for lattice kinematics where energy is injected at the weak vertex.

\section{Hadronic Parity Violating Interactions}
\label{sec:PVO}

The four-quark operators that contribute to flavor-conserving, low-energy 
PV processes can be classified by how they act in 
isospin space, $\Delta I=0,1,2$.
Their
QCD evolution from the weak-scale down to the strong interaction scale
has been computed previously~\cite{DDH,renorm,DSLS}.
In this work we focus on the $\Delta I=1$ interaction, as it is only 
in this channel that the LO operator in the chiral expansion is
momentum independent.
In QCD, the effective Lagrange density for $\Delta I=1$ interactions at the
quark-level is~\cite{DSLS,KS93}, including strange-quark operators
\begin{eqnarray}
{\cal L}^{\Delta I=1} & = & 
-{G_F\over\sqrt{2}}\ {\sin^2\theta_w\over 3}
\sum_i\ 
\left[\ C_i^{(1)}(\mu)\ \theta_i(\mu)\ +\ 
S_i^{(1)}(\mu)\ \theta_i^{(s)}(\mu)
\ \right]
\ \ \ ,
\label{eq:PVq}
\end{eqnarray}
where $\theta_w$ is the weak mixing angle, and 
where the four-quark operators are
\begin{eqnarray}
\theta_1 & = & 
\overline{q}^\alpha\gamma^\mu q_\alpha\ \overline{q}^\beta\gamma_\mu\gamma_5
\tau^3 q_\beta
\qquad , \qquad 
\theta_2 \ = \ 
\overline{q}^\alpha\gamma^\mu q_\beta\ \overline{q}^\beta\gamma_\mu\gamma_5
\tau^3 q_\alpha
\nonumber\\
\theta_3 & = & 
\overline{q}^\alpha\gamma^\mu \gamma_5 q_\alpha\ \overline{q}^\beta\gamma_\mu
\tau^3 q_\beta
\qquad , \qquad 
\theta_4 \ = \ 
\overline{q}^\alpha\gamma^\mu \gamma_5 q_\beta\ \overline{q}^\beta\gamma_\mu
\tau^3 q_\alpha
\nonumber\\
\theta_1^{(s)} & = & 
\overline{s}^\alpha\gamma^\mu s_\alpha\ \overline{q}^\beta\gamma_\mu\gamma_5
\tau^3 q_\beta
\qquad , \qquad 
\theta_2^{(s)} \ = \ 
\overline{s}^\alpha\gamma^\mu s_\beta\ \overline{q}^\beta\gamma_\mu\gamma_5
\tau^3 q_\alpha
\nonumber\\
\theta_3^{(s)} & = & 
\overline{s}^\alpha\gamma^\mu \gamma_5 s_\alpha\ \overline{q}^\beta\gamma_\mu
\tau^3 q_\beta
\qquad , \qquad 
\theta_4^{(s)} \ = \ 
\overline{s}^\alpha\gamma^\mu \gamma_5 s_\beta\ \overline{q}^\beta\gamma_\mu
\tau^3 q_\alpha
\ \ \ ,
\label{eq:fops}
\end{eqnarray}
and their coefficients at the chiral symmetry breaking scale 
are~\cite{DSLS,KS93}
\begin{eqnarray}
C^{(1)}(\Lambda_\chi) & = & 
\left(\ 1.10\  ,\  0.068\  ,\  0.234\  ,\  -0.697\ \right)
\nonumber\\
S^{(1)}(\Lambda_\chi)  & = & 
\left(\ 5.61\ ,\ -1.90\ ,\ 4.74\ ,\ -2.67\ \right)
\ \ \ ,
\label{eq:cvals}
\end{eqnarray}
where we have neglected inter-generational mixing through the Cabibbo angle.
In the limit of vanishing $\theta_w$  and degenerate quark 
weak-isospin doublets, 
the standard model possesses a global symmetry that forbids $\Delta I=1,2$
parity-violation~\cite{DSLS}. 
This symmetry is reflected in the relative magnitudes of the 
$S^{(1)}$ versus the $C^{(1)}$.
If strange quarks were to play no role in the structure of the nucleon, 
two-flavor lattice QCD simulations would provide an accurate calculation
of the matrix elements of $\theta_{1,..4}$, 
but to the extent to which strange matrix
elements are non-zero, the two-flavor simulations  provide only 
part of the PV matrix elements.
The role that strange quarks may play in these matrix elements has been
considered in various unjustified approaches, e.g. Ref.~\cite{DDH}, 
\cite{DSLS} and \cite{Dstrange}, and 
the strange quark is found to make a sizable contribution~\cite{Dstrange}
to the PV momentum independent interaction, $h_{\pi NN}^{(1)}$. 
However, the computation of these matrix elements in two-flavor 
simulations will be a vital 
first-step towards a rigorous understanding of flavor conserving, hadronic
parity-violation.

In order to match onto a chiral Lagrange density that describes low-energy PV
processes, it is convenient to introduce the objects
$X^a_{L,R}$, defined as~\cite{KS93}
\begin{eqnarray}
X_L^a & = &   \xi^\dagger\ \tau^a\ \xi
\ \ ,\ \ 
X_R^a \ =\ \xi\ \tau^a\ \xi^\dagger
\ \ \ ,
\label{eq:theXs}
\end{eqnarray}
where
\begin{eqnarray}
\xi & = & \exp\left({i\ M\over f}\right)
\ \ \ ,\ \ \ 
M \ =\  \left(\matrix{ \pi^0/\sqrt{2} &\pi^+ \cr \pi^- 
& -\pi^0/\sqrt{2}}\right)
\ \ \ ,
\label{eq:phidefmod}
\end{eqnarray}
and where $\xi\rightarrow L \xi  U^\dagger= U\xi R^\dagger$, and 
$X_{L,R}^a\rightarrow U X_{L,R}^a U^\dagger$ 
under chiral transformations. In our convention, the pion decay constant,
$f=132~{\rm MeV}$.
At LO in the chiral expansion, 
$\Delta I=1$ PV couplings between the 
pions, the nucleons and the $\Delta$'s 
are described by the  Lagrange 
density~\cite{KS93}~\footnote{Ref.~\cite{ZPHRM} uses similar notation 
with $h_{\pi NN}^{(1)}=h_\pi$
and  $h_{\pi \Delta\Delta}^{(1)}=h_\Delta$.},
\begin{eqnarray}
& & {\cal L}_{\rm wk} \ =\  -h_{\pi NN}^{(1)}\  {f\over 4} \ 
\overline{N} \left[\ X_L^3\ -\ X_R^3\ \right] N
\ -\ 
h_{\pi \Delta\Delta}^{(1)}\  {f\over 4} \
\overline{T}^{abc,\mu}
\left[\ X_L^3\ -\ X_R^3\ \right]^d_c T_{abd,\mu}
\nonumber\\
 & & \rightarrow  
i \pi^-  \left[h_{\pi NN}^{(1)}  \overline{n} p 
+
{h_{\pi \Delta\Delta}^{(1)}\over\sqrt{3}}
\overline{\Delta}^{+\mu}\Delta^{++}_\mu
+
{2 h_{\pi \Delta\Delta}^{(1)} \over 3}\overline{\Delta}^{0\mu}\Delta^{+}_\mu
+
{ h_{\pi \Delta\Delta}^{(1)} \over\sqrt{3}}
\overline{\Delta}^{-\mu}\Delta^{0}_\mu
\right]+ {\rm h.c.}
\ .
\label{eq:weaktree}
\end{eqnarray}

\section{Extension to Partially-Quenched QCD}
\label{sec:PVPQCD}

The extension of the PV flavor-conserving weak operators from QCD to PQQCD
is analogous to the extension of strangeness-changing operators,
as discussed in detail in Ref.~\cite{GP01a}.
It is convenient to rewrite the operators in eq.~(\ref{eq:fops})
in terms of operators with well-defined 
properties under chiral transformations. Neglecting the strange-quark
operators one finds,

\begin{eqnarray}
\tilde\theta_1 & = & 
\overline{q}^\alpha \tau^3 \gamma^\mu (1-\gamma_5) q_\alpha\ 
\overline{q}^\beta \gamma_\mu (1-\gamma_5) q_\beta
\ \ ,\qquad
\tilde\theta_2\ =\  
\overline{q}^\alpha \tau^3 \gamma^\mu (1+\gamma_5) q_\alpha\ 
\overline{q}^\beta \gamma_\mu (1-\gamma_5) q_\beta
\nonumber\\
\tilde\theta_3 & = & 
\overline{q}^\alpha \tau^3 \gamma^\mu (1-\gamma_5) q_\alpha\ 
\overline{q}^\beta \gamma_\mu (1+\gamma_5) q_\beta
\ \  ,\qquad
\tilde\theta_4\ =\  
\overline{q}^\alpha \tau^3 \gamma^\mu (1+\gamma_5) q_\alpha\ 
\overline{q}^\beta \gamma_\mu (1+\gamma_5) q_\beta
\ ,
\label{eq:wkch}
\end{eqnarray}
and operators $\tilde\theta_5$..$\tilde\theta_8$, corresponding to the 
other color-contraction.
These objects transform as 
$({\bf 3},{\bf 1})$, $({\bf 1},{\bf 3})$,
$({\bf 3},{\bf 1})$, $({\bf 1},{\bf 3})$, respectively under $SU(2)_L\otimes
SU(2)_R$ chiral transformations.
The most general extension 
of this operator set to PQQCD
is quite messy, as one ends up dealing with objects of the form
$({\bf A}, {\bf A})$ of $SU(4|2)_L\otimes SU(4|2)_R$, where ${\bf A}$ denotes
the adjoint representation of $SU(4|2)$.
Therefore, for simplicity reasons alone, we will only consider
extensions of the form $({\bf A}, {\bf 1})$ and 
$({\bf 1}, {\bf A})$~\footnote{The flavor structure can be 
further extended when lattice simulations require.}.
In this case, we can
define the weak flavor matrix
\begin{eqnarray}
\overline{\tau}^3 & = & 
\left(\ 1\  ,\  -1\ ,\  h_j \ ,\  h_l\ ,\   h_j \ ,\  h_l\ \right)
\ \ \ ,
\label{eq:wkcharge}
\end{eqnarray}
where $h_j$ and $h_l$ are arbitrary weak charges.
This matrix is supertraceless and its matrix elements reduce to those of QCD
when the sea-quark ($j$ and $l$) masses become degenerate with the 
valence-quark masses.
The quark mass matrix in PQQCD is given by 
$m_Q={\rm diag}(m_u,m_d,m_j,m_l,m_u,m_d)$.
Further, in analogy with QCD~\cite{KS93}  we define
\begin{eqnarray}
^{(PQ)} X_L^a & = & \xi^\dagger \ \overline{\tau}^a \ \xi
\ \ ,\ \ 
^{(PQ)} X_R^a \ = \ \xi \ \overline{\tau}^a \ \xi^\dagger
\ \ \ .
\label{eq:PQXmats}
\end{eqnarray}
The LO Lagrange density describing the low-energy PV interactions
in PQ$\chi$PT is
\begin{eqnarray}
{\cal L}_{\rm wk} & = & 
-W_1 {f\over 4} \ \left(\overline{\cal B} {\cal B} 
\left[\ ^{(PQ)} X_L^3\ -\ ^{(PQ)} X_R^3\ \right] \right)
\ -\ 
W_2\ {f\over 4}\left(\overline{\cal B} 
\left[\ ^{(PQ)} X_L^3 - ^{(PQ)} X_R^3\ \right]
{\cal B} \right)
\nonumber\\
& & -  
W_3 {f\over 4}\left(\overline{\cal T}^\nu 
\left[\ ^{(PQ)} X_L^3 -^{(PQ)}  X_R^3\ \right]
{\cal T}_\nu \right)
\ ,
\label{eq:wkints}
\end{eqnarray}
where ${\cal B}$, which contains the nucleons, and ${\cal T}_\nu$, which
contains the $\Delta$-resonances,
transform in the ${\bf 70}$ and ${\bf 44}$ representations of
$SU(4|2)$, respectively~\cite{BS02b}. The contraction
of flavor indices, $\left( \ \right)$, can be found in Ref.~\cite{LS96}.
At tree-level one can make the identification

\begin{eqnarray}
h_{\pi NN}^{(1)} & = & {1\over 3}\left( 2W_1 - W_2\ \right)
\ \ ,\ \ 
h_{\pi \Delta\Delta}^{(1)}\ =\ W_3
\ \ .
\end{eqnarray}
The LO weak coupling between the nucleons and $\Delta$-resonances
involves one derivative and hence is higher order in the chiral expansion.

The Lagrange density describing the parity-conserving interactions of the 
${\bf 70}$ and ${\bf 44}$ with the
pseudo-Goldstone bosons at LO in the chiral expansion
is~\cite{LS96},
\begin{eqnarray}
{\cal L} & = & 
2\alpha\ \left(\overline{\cal B} S^\mu {\cal B} A_\mu\right)
\ +\ 
2\beta\ \left(\overline{\cal B} S^\mu A_\mu {\cal B} \right)
\ +\  
2{\cal H} \left(\overline{\cal T}^\nu S^\mu A_\mu {\cal T}_\nu \right)
\nonumber\\
& &  
\ +\ 
\sqrt{3\over 2}{\cal C} 
\left[\ 
\left( \overline{\cal T}^\nu A_\nu {\cal B}\right)\ +\ 
\left(\overline{\cal B} A^\nu {\cal T}_\nu\right)\ \right]
\ ,
\label{eq:ints}
\end{eqnarray}
where $S^\mu$ is the covariant spin-vector~\cite{JMheavy,JMaxial,Jmass},
and $A^\mu \ =\  {i\over 2}\left(\ \xi\partial^\mu\xi^\dagger
\ - \ 
\xi^\dagger\partial^\mu\xi \ \right)$.
A comparison with the LO interaction Lagrange density of QCD
yields the tree-level identification~\cite{BS02b} 

\begin{eqnarray}
\alpha & = & {4\over 3} g_A\ +\ {1\over 3} g_1
\ \ \ ,\ \ \ 
\beta \ =\ {2\over 3} g_1 - {1\over 3} g_A
\ \ \ ,\ \ \ 
{\cal H} \ =\ g_{\Delta\Delta}
\ \ \ ,\ \ \ 
{\cal C} \ =\ -g_{\Delta N}
\ \ \ ,
\label{eq:axrels}
\end{eqnarray}
where $g_A$, $g_{\Delta\Delta}$ and $g_{\Delta N}$ are the $NN$, $\Delta\Delta$
and $\Delta N$ isovector axial couplings, respectively, and 
($g_A$+$g_1$) is the isosinglet axial coupling.
For details about this and other aspects of PQ$\chi$PT, we refer the 
reader to Ref.~\cite{BS02b}.

\section{${\hlittle}_{\pi NN}^{(1)}$ at One Loop with 
$P_\pi^\mu=0$ }
\label{sec:hpinn}

In QCD, the LO pion-nucleon PV interaction is generated by the Lagrange
density in eq.~(\ref{eq:weaktree}).  At higher orders,
the PV pion-nucleon interaction will receive contributions
from one-loop diagrams and also from terms at next-to-LO (NLO) in the chiral
expansion involving a single insertion of $m_q$,
\begin{eqnarray}
{\cal L} & = & - \tilde c_1\  {f\over 4} \ 
\overline{N} \{ {\cal M}^{QCD}_+ , X_L^3-X_R^3\} N
\ -\  \tilde c_2 \ {f\over 4} \ 
\overline{N}
\left( X_L^3-X_R^3\right) N \ {\rm tr}\left({\cal M}^{QCD}_+ \right)
\ \ \ .
\label{eq:weakctsQ}
\end{eqnarray}
Since the pion mass is much greater than the proton-neutron mass difference, 
one
does not have all three particles on their mass-shells.
Therefore, we compute the one-loop corrections to the weak vertex
in both QCD and PQQCD.  
The one-loop calculation of the vertex has been performed previously 
in QCD~\cite{ZPHRM}, and
explicit computation in the isospin limit 
and with vanishing pion four-momentum, $P_\pi^\mu$, gives

\begin{eqnarray}
\Gamma_{\overline{n}p\pi^+}
& = & 
-h_{\pi NN}^{(1)}\ -\ 2\overline{m}\ \left(\tilde c_1+\tilde c_2\right) 
\nonumber\\ & &
+ {1\over 16\pi^2 f^2}
\left[\ 
h_{\pi NN}^{(1)}
\left({2\over 3}+{3\over 2} g_A^2\right)L_\pi
\ +\ 
h_{\pi\Delta\Delta}^{(1)} {20\over 9}   g_{\Delta N}^2 J_\pi
\ \right]
\label{eq:QCDweakV}
\ \ \ ,
\end{eqnarray}
where ${\overline m}=m_u=m_d$ in the isospin limit.
The loop functions are defined as
${L_\pi} = m_\pi^2\log\left({m_\pi^2/\mu^2}\right)$, and 
$J_\pi=J(m_\pi,\Delta,\mu)$ with

\begin{eqnarray}
J(m,\Delta,\mu) & = & 
\left(m^2-2\Delta^2\right)\log\left({m^2\over\mu^2}\right)
+2\Delta\sqrt{\Delta^2-m^2}
\log\left({\Delta-\sqrt{\Delta^2-m^2+ i \epsilon}\over
\Delta+\sqrt{\Delta^2-m^2+ i \epsilon}}\right)
\ \ \ ,
\label{eq:decfun}
\end{eqnarray}
and $\Delta$ is the $\Delta$-nucleon mass splitting.
Our results agree with those of Ref.~\cite{ZPHRM}, 
once differences in conventions are accounted for.

\begin{figure}[!ht]
\centerline{{\epsfxsize=3.0in \epsfbox{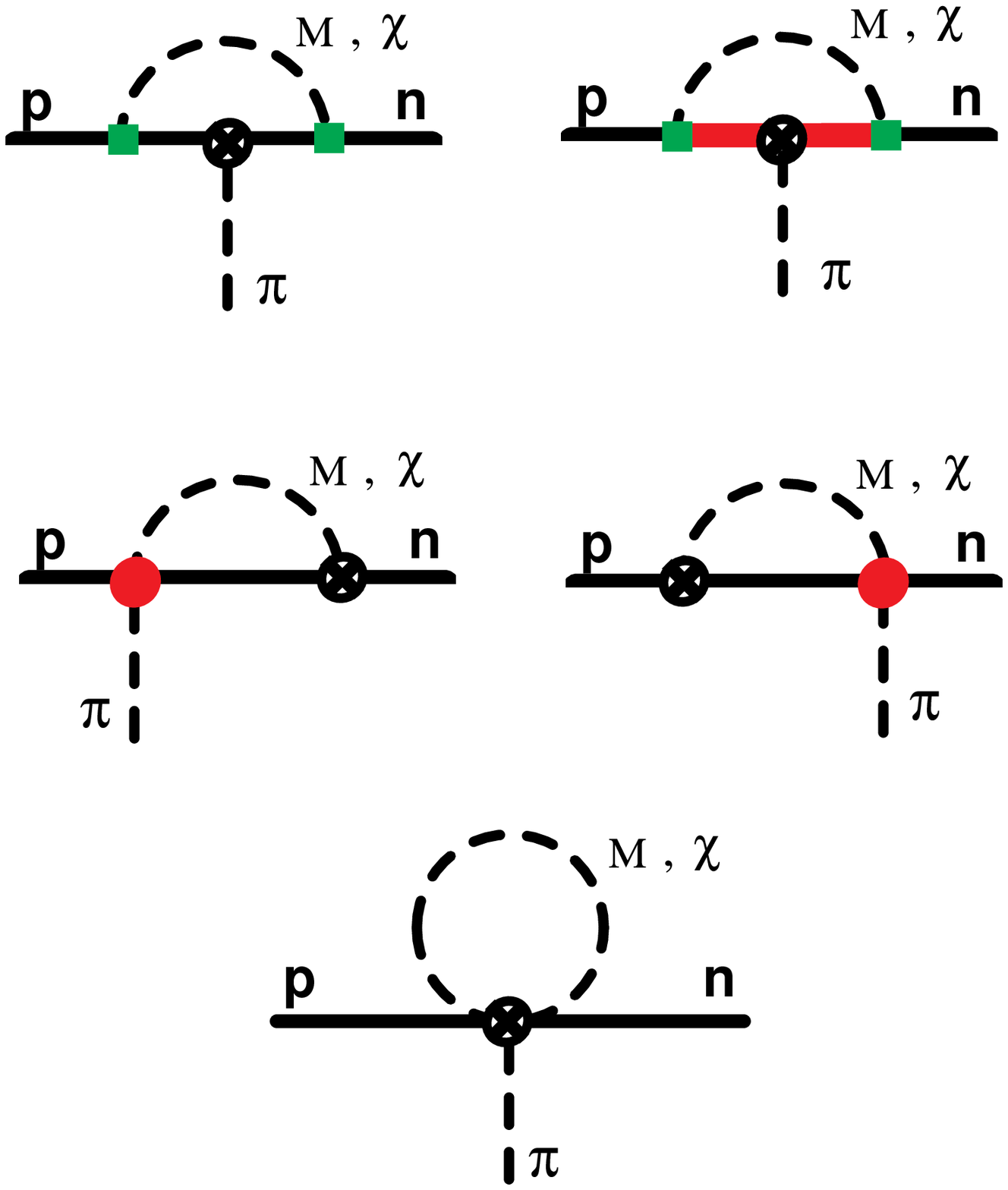}}} 
\vskip 0.15in
\noindent
\caption{\it 
One-loop graphs that give contributions of the form 
$\sim m_Q \log m_Q$ to the momentum independent 
parity-violating interaction $h_{\pi NN}^{(1)}$.
A solid, thick-solid and dashed line denote a
{\bf 70}-nucleon, {\bf 44}-resonance, and a meson, respectively.
The solid-squares denote an axial coupling given in eq.(\ref{eq:ints}),
while the crossed circle denotes an insertion of the parity-violating
pion-nucleon  operators with coefficients $W_{1,2,3}$ in eq.~(\ref{eq:wkints}).
The solid circle denotes an insertion of the strong two-pion vertex from the 
nucleon kinetic energy term.
}
\label{fig:hpi}
\vskip .2in
\end{figure}

In PQQCD the counterterms involving a single insertion of $m_Q$
that contribute to the PV pion-nucleon interaction at the 
one-loop level are 
\begin{eqnarray}
{\cal L}^{(ct)}
& = &
-{f\over 2}
\left[\ 
c_1\  \cbb^{kji}\ \{\  
^{(PQ)}X_L^3-^{(PQ)}X_R^3\ ,\ 
{\cal M}_+\ \}^n_i\ \cb_{njk}
\right.\nonumber\\ & & \left.
+\ 
c_2\ (-)^{(\eta_i+\eta_j)(\eta_k+\eta_n)}\ 
\cbb^{kji}\ \{\ ^{(PQ)}X_L^3-^{(PQ)}X_R^3\ ,\ {\cal M}_+\ \}^n_k\ \cb_{ijn}
\right.\nonumber\\ & & \left.
+\ 
c_3\  (-)^{\eta_l (\eta_j+\eta_n)}\
\cbb^{kji}\  \left(^{(PQ)}X_L^3-^{(PQ)}X_R^3\right)^l_i\ 
\left( {\cal M}_+\right)^n_j
\cb_{lnk}
\right.\nonumber\\ & & \left.
+\ 
c_4 \  (-)^{\eta_l \eta_j}\ 
\cbb^{kji}\ \left(  
\left(^{(PQ)}X_L^3-^{(PQ)}X_R^3\right)^l_i\left( {\cal M}_+\right)^n_j
+\left( {\cal M}_+\right)^l_i
 \left(^{(PQ)}X_L^3-^{(PQ)}X_R^3\right)^n_j \right)
\cb_{nlk}
\right.\nonumber\\ & & \left.
+\ c_5\  (-)^{\eta_i(\eta_l+\eta_j)}\ 
\cbb^{kji} \left(^{(PQ)}X_L^3-^{(PQ)}X_R^3\right)^l_j 
\left( {\cal M}_+\right)^n_i
\cb_{nlk}
\right.\nonumber\\ & & \left.
+\ c_6\  \cbb^{kji}  \left(^{(PQ)}X_L^3-^{(PQ)}X_R^3\right)^l_i \cb_{ljk}
\ {\rm str}\left( {\cal M}_+ \right) 
\right.\nonumber\\ & & \left.
\ +\ c_7\  \ (-)^{(\eta_i+\eta_j)(\eta_k+\eta_n)}\ 
\cbb^{kji}  \left(^{(PQ)}X_L^3-^{(PQ)}X_R^3\right)^n_k \cb_{ijn}
\ {\rm str}\left( {\cal M}_+ \right) 
\right.\nonumber\\ & & \left.
+\ c_8\ \cbb^{kji}\ \cb_{ijk} 
\ {\rm str}\left(\left(^{(PQ)}X_L^3-^{(PQ)}X_R^3\right)\   
{\cal M}_+ \right) \ 
\right]\
\ ,
\label{eq:wkcts}
\end{eqnarray}
where 
${\cal M}_+={1\over 2}\left(\xi^\dagger m_Q\xi^\dagger + \xi m_Q\xi\right)$.
The vertex at one-loop level can be written as 
\begin{eqnarray}
^{(PQ)}\Gamma_{\overline{n}p\pi^+}
& = &
\rho
\ +\ {1\over 16\pi^2 f^2}
\left(\ 
\eta^{0}
\ +\ h_{j}\ \eta^{j}
\ +\ h_l\ \eta^{l}
\ \right)
\ + c^{0} + h_{j}\ c^{j} + h_l\ c^{l}
\ ,
\label{eq:wkmatPQ}
\end{eqnarray}
where the diagrams in Fig.~\ref{fig:hpi}
give
\begin{eqnarray}
\rho & = & 
-{1\over 3}\left( 2 W_1-W_2\ \right)
\nonumber\\
\eta^{0} & = & 
- {\rho\over 6}\left[\ L_{ju}+L_{jd}+L_{lu}+L_{ld} 
+ R_{\eta_u , \eta_u}+ R_{\eta_d , \eta_d}-2 R_{\eta_u , \eta_d}
\ \right]
\ +\ \rho \ 3 g_A^2 R_{\eta_u , \eta_d}\ 
\nonumber\\
& & 
\ +\ {g_1 g_A\over 4}\left[\ 
\left(W_1-2 W_2\right)\left(2 L_{ud}-L_{jd}-L_{ju}-L_{ld}-L_{lu} 
\right)
\right.\nonumber\\
& & \left.\qquad\qquad
+6 \rho\ \left(
R_{\eta_u , \eta_u} + 2 R_{\eta_u , \eta_d} +  R_{\eta_d , \eta_d}
\right)
-\ 3 W_1 \left(L_{uu}+L_{dd}\right)\right]
\nonumber\\
& & 
- {g_1^2\over 8}\left[\ \left(W_1-2 W_2\right)\left( L_{uu}+L_{dd}\right)
\ -\ 6 \rho\ \left(
R_{\eta_u , \eta_u} + 2 R_{\eta_u , \eta_d} +  R_{\eta_d , \eta_d}
\right)
\right.\nonumber\\ & & \left.\qquad
+ 3 W_1 \left( L_{ju}+L_{jd}+ L_{lu}+L_{ld}- 2 L_{ud}\right)
\ \right]
\nonumber\\
&  + &\ g_{\Delta N}^2\ {2\over 9}\ W_3\ \left(\ 
J_{dd} + J_{uu} + 4 J_{ud} + J_{jd} + J_{ju} + J_{ld} + J_{lu}
\right.\nonumber\\ & & \left.\qquad\qquad
+ 2 {\cal T}_{\eta_u , \eta_u} + 2 {\cal T}_{\eta_d , \eta_d} - 4 
{\cal T}_{\eta_u , \eta_d}\ \right)
\nonumber\\
\eta^{j} & = & \eta^{l}  =0
 \nonumber\\
c^0 & = & -\left(m_u+m_d\right) {1\over 3}\left(\ 
-2 c_1 + 4 c_2 - c_3 - c_4 + 2 c_5\ \right)
\ -\ \left(m_j+m_l\right) {2\over 3}\left(\ 
- c_6 + 2 c_7\ \right)
  \nonumber\\
c^j & = & c^l\ =\ 0
\ \ \ ,
\label{eq:wkmatmore}
\end{eqnarray}
where $R_{x , y} = {\cal H}(L_x, L_y, L_X)$ and
${\cal T}_{x , y} = {\cal H}(J_x, J_y, J_X)$, with

\begin{eqnarray}
{\cal H}_{ab}( A, B, C) \ &=&\  
-{1\over 2}\left[\ 
{(m_{jj}^2-m_{\eta_a}^2)(m_{ll}^2-m_{\eta_a}^2)\over 
(m_{\eta_a}^2-m_{\eta_b}^2)(m_{\eta_a}^2-m_X^2)}\  A
-
{(m_{jj}^2-m_{\eta_b}^2)(m_{ll}^2-m_{\eta_b}^2)\over 
(m_{\eta_a}^2-m_{\eta_b}^2)(m_{\eta_b}^2-m_X^2)}\  B
\right.\nonumber\\ & & \left.\qquad
\ +\ 
{(m_X^2-m_{jj}^2)(m_X^2-m_{ll}^2)\over 
(m_X^2-m_{\eta_a}^2)(m_X^2-m_{\eta_b}^2)}\  C
\ \right]
\ \ \ ,
\label{eq:HPsdef}
\end{eqnarray}
where the mass, $m_X$, is given by 
$m_X^2 = {1\over 2}\left( m_{jj}^2+ m_{ll}^2 \right)$.
The expression in eq.~(\ref{eq:wkmatmore}) collapses down to the 
isospin-symmetric
QCD expression given in 
eq.~(\ref{eq:QCDweakV}) in the limit,
$m_j,m_l,m_u,m_d\rightarrow\overline{m}$.
Wavefunction renormalization can be performed for 
those particles on their mass-shell using
(expressions for $w_p$ and $w_n$ are given in Ref.~\cite{BS02b}
and $w_\pi$ in the isospin limit is given in Ref.~\cite{Pqqcd}),
\begin{eqnarray}
w_\pi & = & {1\over 3}\left[\ 
-L_{ju}-L_{jd}-L_{lu}-L_{ld} + 2 R_{\eta_u , \eta_d} 
- R_{\eta_u , \eta_u} - R_{\eta_d , \eta_d} \ \right]
\nonumber\\
w_p & = & g_A^2\left( L_{ud}+L_{uu}+2 L_{ju}+2 L_{lu} + 3 R_{\eta_u , \eta_u}
\right)
\nonumber\\
& & 
\ +\ g_1 g_A \left( 2 L_{uu} - L_{ud} + L_{ju} + L_{lu} 
+ 3 R_{\eta_u , \eta_u} + 3 R_{\eta_u , \eta_d}
\right)
\nonumber\\
& & 
\ +\ 
{g_1^2\over 4} \left( L_{uu} - 5 L_{ud} + 2 L_{lu} + 3 L_{ld} + 2 L_{ju} 
+ 3 L_{jd} + 3 R_{\eta_u , \eta_u} + 6 R_{\eta_u , \eta_d}
+ 3 R_{\eta_d , \eta_d}
\right)
\nonumber\\
& & 
+{1\over 3} g_{\Delta N}^2 \left( 5 J_{ud} + J_{uu} + J_{ju}+J_{lu}+2 J_{jd}+2
  J_{ld} + 2 {\cal T}_{\eta_u , \eta_u} + 2 {\cal T}_{\eta_d , \eta_d} 
-4  {\cal T}_{\eta_u , \eta_d} \right)
\nonumber\\
w_n & = & g_A^2\left( L_{dd}+L_{ud}+2 L_{jd}+2 L_{ld} + 3 R_{\eta_d , \eta_d}
\right)
\nonumber\\
& & 
+\ g_1 g_A\left( 2 L_{dd} - L_{ud} + L_{jd} + L_{ld} + 
3 R_{\eta_u , \eta_d} + 3 R_{\eta_d , \eta_d}
\right)
\nonumber\\
& & 
+ {g_1^2\over 4}\left( 
L_{dd} - 5 L_{ud} + 2 L_{jd} + 3 L_{ju} + 2 L_{ld} + 3 L_{lu} 
+ 3 R_{\eta_u , \eta_u} + 6 R_{\eta_u , \eta_d} + 3 R_{\eta_d , \eta_d}
\right)
\nonumber\\
& & 
+{1\over 3} g_{\Delta N}^2 \left( 5 J_{ud} + J_{dd} + 2 J_{ju}+2J_{lu}
+ J_{jd}+  J_{ld} + 2 {\cal T}_{\eta_u , \eta_u} 
+ 2 {\cal T}_{\eta_d , \eta_d} -4  {\cal T}_{\eta_u , \eta_d} \right)
\ \ \ ,
\label{eq:wpi}
\end{eqnarray}
by adding a contribution
\begin{eqnarray}
\delta ^{(PQ)}\Gamma_j & = & -{1\over 16\pi^2 f^2} \ \rho\  w_j
\ \ \ .
\end{eqnarray}

\section{${\hlittle}_{\pi NN}^{(1)}$ at One Loop with 
Lattice Kinematics }
\label{sec:hpinnlat}

The analysis of the previous section will not be as useful 
in determining $h_{\pi NN}^{(1)}$ from the lattice 
as one would naively assume~\footnote{We are indebted to Steve Sharpe
for making this point clear to us.}.
While the $P_\pi^\mu=0$ limit is natural to use from the viewpoint of
a momentum and $m_q$ expansion, the fact that lattice simulations only
measure on-shell to on-shell amplitudes means that the pions and nucleons are
on their mass-shells in both the initial and final states.
The extraction of $h_{\pi NN}^{(1)}$ from $N\rightarrow N\pi$
requires an injection of energy at the PV weak vertex
which can occur because  the weak operator is inserted on one time-slice only.
Therefore, we must include contributions from operators that are total
derivatives, which usually vanish.
Recently, 
chiral perturbation theory has been used to describe $K\rightarrow \pi\pi$
with the  kinematics appropriate for a lattice determination of the matrix
elements of the relevant four-quark operators,
$m_K^{\rm latt}=m_\pi^{\rm latt}$ and $m_K^{\rm latt}=2 m_\pi^{\rm latt}$,
including the necessary total derivative terms~\cite{kpipi}.
In QCD, the LO Lagrange density describing PV interactions 
is given in eq.~(\ref{eq:weaktree}), while the Lagrange density at NLO is
\begin{eqnarray}
& & {\cal L}_{\rm wk}^{(NLO)} \ =\  
h_{D}^{(1)}\  {1\over 4} \ 
i v\cdot D\ \overline{N} \left[\ X_L^3\ -\ X_R^3\ \right] N
\ \ \ ,
\label{eq:NLOweak}
\end{eqnarray}
where $v^\mu$ is the nucleon four-velocity.  This is the leading contribution
from a heavy baryon reduction of 
$i D^\mu \overline{N} \gamma_\mu \left[\ X_L^3\ -\ X_R^3\ \right] N$.
Given baryon number conservation, the total derivative gives a non-zero
contribution from the energy and momentum injected by the 
$X_L^3\ -\ X_R^3$ insertion.
Working in the frame where the initial state nucleon (proton) is at rest,
$v^\mu = (1,0,0,0)$,
the amplitude at NLO resulting from eq.~(\ref{eq:weaktree}) and 
eq.~(\ref{eq:NLOweak}) is 
\begin{eqnarray}
{\cal A}_{\overline{n}p\pi}\ =\ 
\langle n\pi | i \int d^3 {\bf x}\  {\cal L}^{\Delta I=1} (E) 
| p\rangle
& = & -\overline{U}_n \ 
\left[\ h_{\pi NN}^{(1)}\ +\ h_{D}^{(1)} {E\over f}\ \right]
U_p
\ \ \ .
\label{eq:NLOtree}
\end{eqnarray}
where $E$ is the energy injected by the weak vertex.
In order to produce an on-shell $n\pi$ final state, the injected energy must
exceed $E\ge m_\pi + M_n-M_p$.
Near threshold, where the final state neutron and pion are at rest and 
$E =  m_\pi + M_n-M_p$, 
the contribution from the total-derivative operator, $h_{D}^{(1)}$,
scales as $\sim m_q^{1/2}$, and is formally dominant over the 
loop corrections and counterterms of the previous section.

\begin{figure}[!ht]
\centerline{{\epsfxsize=3.5in \epsfbox{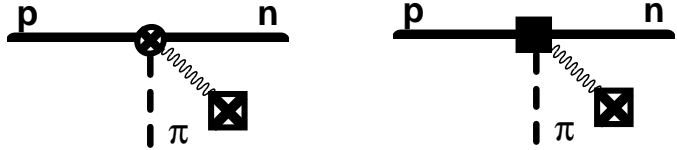}}} 
\vskip 0.15in
\noindent
\caption{\it 
Tree-level contributions to the parity-violating vertex with lattice
kinematics. 
The crossed circle denotes an insertion of the parity-violating
pion-nucleon operators with coefficients $W_{1,2}$ in eq.~(\ref{eq:wkints}).
The solid square denotes an insertion of the energy-momentum dependent 
operators with coefficients $W_{D1, D2}$ in eq.~(\ref{eq:NLOwkints}).
The crossed-box denotes an insertion of energy-momentum at the weak vertex.
}
\label{fig:hpilatttree}
\vskip .2in
\end{figure}
\begin{figure}[!ht]
\centerline{{\epsfxsize=3.0in \epsfbox{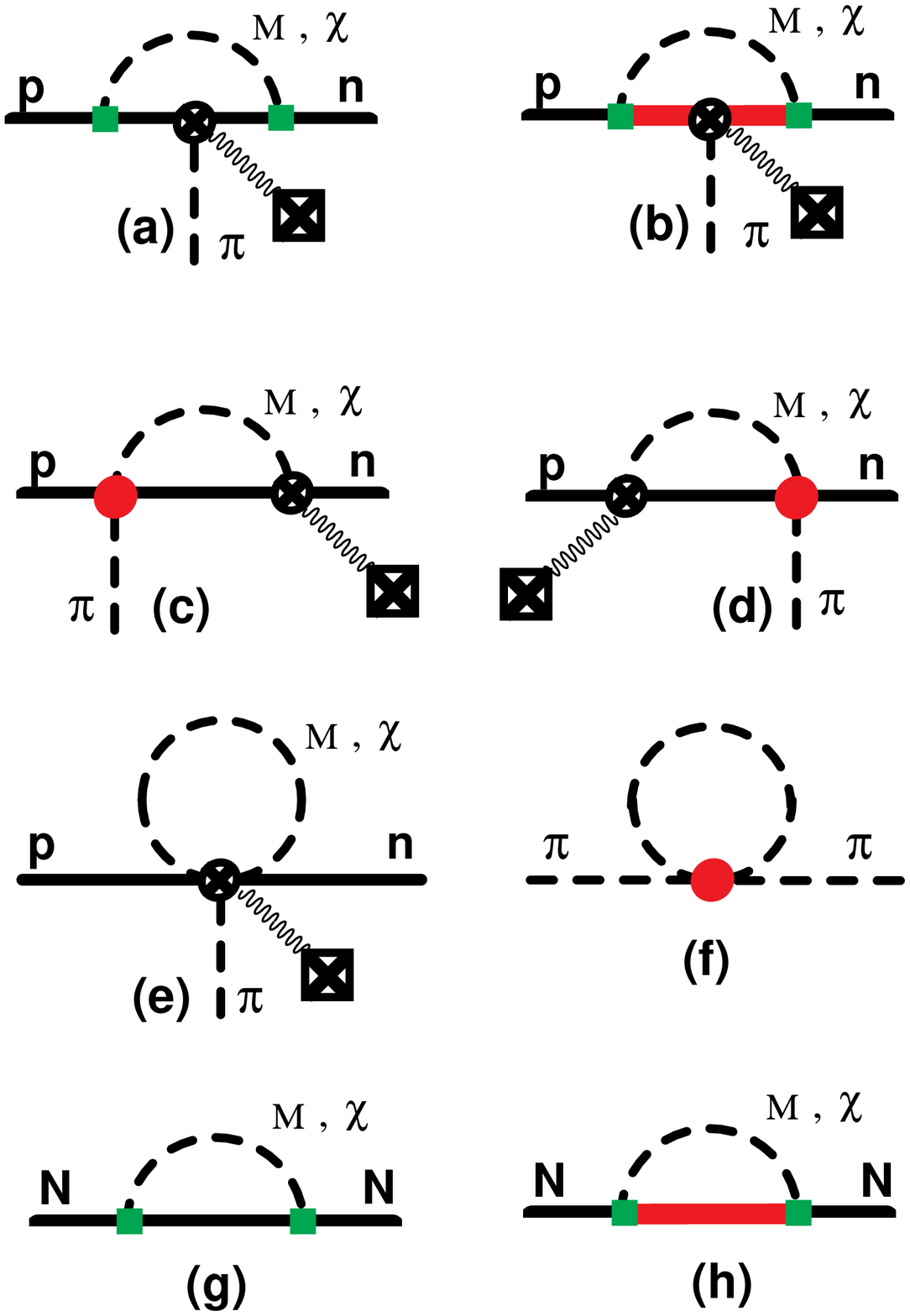}}} 
\vskip 0.15in
\noindent
\caption{\it 
One-loop graphs that give contributions of the form 
$\sim m_Q \log m_Q$ to the momentum independent 
parity-violating interaction $h_{\pi NN}^{(1)}$.
A solid, thick-solid and dashed line denote a
{\bf 70}-nucleon, {\bf 44}-resonance, and a meson, respectively.
The small solid squares denote an axial coupling given in eq.(\ref{eq:ints}),
while the crossed circle denotes an insertion of the parity-violating
pion-nucleon  operators with coefficients $W_{1,2,3}$ in eq.~(\ref{eq:wkints}).
The solid circle denotes an insertion of the strong two-pion vertex from the 
nucleon kinetic energy term.
The crossed-box denotes an insertion of energy-momentum at the weak vertex.
Diagrams (a) to (e) contribute to  
vertex renormalization while diagrams (f) to (h)
contribute to  wavefunction renormalization.
}
\label{fig:hpilatt}
\vskip .2in
\end{figure}

In PQQCD, the lagrange density describing PV interactions at NLO is
\begin{eqnarray}
{\cal L}_{\rm wk}^{(NLO)} & = & 
-i {W_{D1}\over 4}\  v\cdot D\ \ \left(\overline{\cal B} {\cal B} 
\left[\ ^{(PQ)} X_L^3\ -\ ^{(PQ)} X_R^3\ \right] \right)
\nonumber\\
& & \ -\ 
i {W_{D2}\over 4}\ v\cdot D\  \left(\overline{\cal B} 
\left[\ ^{(PQ)} X_L^3 - ^{(PQ)} X_R^3\ \right]
{\cal B} \right)
\ .
\label{eq:NLOwkints}
\end{eqnarray}
The one-loop calculation of the previous section, with minor modifications,
provides the leading non-analytic contributions arising at N$^2$LO in the
chiral expansion.  All other contributions are formally suppressed in the
chiral limit. 
The matrix element at one-loop can be written as, keeping only leading
non-analytic terms,
\begin{eqnarray}
^{(PQ)}{\cal A}_{\overline{n}p\pi^+}
& = & \overline{U}_n \left[\ 
\overline{\rho}
\ +\ {1\over 16\pi^2 f^2}
\left(\ 
\overline{\eta}^{0}
\ -\ {1\over 2}\overline{\rho}\left[\ w_p + w_n + w_\pi\ \right] 
\ +\ h_{j}\ \overline{\eta}^{j}
\ +\ h_l\ \overline{\eta}^{l}
\ \right)
\right]\ U_p
\ ,
\label{eq:wkmatPQlat}
\end{eqnarray}
where the wavefunction renormalization contributions, $w_i$, can be found in
eq.~(\ref{eq:wpi}).
The diagrams in Fig.~\ref{fig:hpilatt}
give, retaining only the leading non-analytic contributions and working 
in the frame where the initial state proton is at rest,
\begin{eqnarray}
\overline{\rho} & = & \rho\ -\ 
{E\over 3 f}\left( 2 W_{D1} - W_{D2}\right)
\nonumber\\
\overline{\eta}^{0} & = & \eta^0\ +\ 
{\rho\over 4}\left[\ \tilde L^{(-)}_{ud} - \tilde L^{(-)}_{dd}
+\tilde L^{(-)}_{jd}+\tilde L^{(-)}_{ld}
+\tilde L^{(+)}_{ud}-\tilde L^{(+)}_{uu}+\tilde L^{(+)}_{ju}
+\tilde L^{(+)}_{lu}\ \right]
\nonumber\\
\overline{\eta}^{j} & = & \eta^j\ +\ 
{\rho\over 4}\left[\ \tilde L^{(+)}_{uu} - \tilde L^{(+)}_{ju}
- \tilde L^{(-)}_{ud} + \tilde L^{(-)}_{jd}
\ \right]
\nonumber\\
\overline{\eta}^{l} & = & \eta^l\ +\ 
{\rho\over 4}\left[\ \tilde L^{(+)}_{ud} - \tilde L^{(+)}_{lu}
- \tilde L^{(-)}_{dd} + \tilde L^{(-)}_{ld}
\ \right]
\ \ \ \ ,
\label{eq:lattloop}
\end{eqnarray}
where $\rho$, $\eta^0$, $\eta^j$ and $\eta^l$ are given in 
eq.~(\ref{eq:wkmatmore}).
The functions 
$\tilde L^{(\pm)}_{ij} = \tilde L (m_{ij},\pm E, \mu)$ are
\begin{eqnarray}
\tilde L (m, +E, \mu) & = & -4 E \left[\ 
E \log\left({m^2\over\mu^2}\right)\ -\ 
\sqrt{E^2-m^2}\ \log\left({-E - \sqrt{E^2-m^2 + i\epsilon}\over
-E + \sqrt{E^2-m^2 + i\epsilon}}\right)
\ \right]
\ \ \ .
\end{eqnarray}
Note that these functions are enhanced by a chiral logarithm compared with
contributions from local counterterms in the chiral limit.
When $E=\pm m_\pi$, 
corresponding to the production of a nucleon and pion at
rest, $\tilde L (m,\pm m, \mu)=-4 m^2 \log\left({m^2/\mu^2}\right)$.
The additional non-analytic contributions in eq.~(\ref{eq:lattloop}) 
result from modifications to diagrams (c) and (d) in Fig.~\ref{fig:hpilatt},
with the other diagrams unchanged.

In the limit where $m_{j,l,u,d}\rightarrow \overline{m}$, this matrix element 
reduces down to that of QCD,
\begin{eqnarray}
& & {\cal A}_{\overline{n}p\pi^+}
\ =\  \overline{U}_n \left[\ -h_{\pi NN}^{(1)}\ -\ h_D^{(1)} {E\over f}
\right.\nonumber\\ & & \left.
\ +\ {1\over 16\pi^2 f^2}\left(\ 
\left(6 g_A^2 + 4 g_{\Delta N}^2 \right) h_{\pi NN}^{(1)} L_\pi
\ +\ {20\over 9} g_{\Delta N}^2 h_{\pi \Delta\Delta}^{(1)} J_\pi
- {1\over 2}\left( \tilde L^{(+)}_\pi +  \tilde L^{(-)}_\pi \right)
h_{\pi NN}^{(1)}
\ \right)
\right] U_p
\ .
\end{eqnarray}

We have not given the numerous counterterms that are expected to  
appear at N$^2$LO whose $\mu$-dependence will exactly cancel that of the 
one-loop expressions in eq.~(\ref{eq:wkmatPQlat}).
The counterterms are expected to make contributions that are smaller than 
those of the one-loop graphs when a renormalization scale of
$\mu\sim\Lambda_\chi$ is chosen, where $\Lambda_\chi$ is the chiral symmetry
breaking scale.

For an energy injection of 
$E > m_\pi + M_n - M_p$, the amplitude develops an imaginary
part due to the rescattering of the pion in the intermediate state 
(final-state interactions).
Therefore, in accordance with the 
Maiani-Testa theorem~\cite{MT90},  
the simulations must be done at threshold, with $E = m_\pi + M_n - M_p$.

\section{The Anapole Moment and Form Factor of the Nucleon}
\label{sec:anupole}

An object that plays an important role in PV 
$eH$ scattering is the anapole moment of $H$, where $H$ denotes
a generic hadron.
The dominant PV contributions to $eH$ scattering arise
not only from tree-level $Z^0$-exchange between 
$H$ and the electron, 
but also from the exchange of a photon 
along with hadronic PV from the 
four-quark operators in eq.~(\ref{eq:fops}).
For the proton,  
there is a contribution to the matrix element of the 
electromagnetic current in the presence
of hadronic PV interactions, of the form
\begin{eqnarray}
\langle p | j^\mu_{\rm em} (q) | p \rangle
& = & {2\over M_p^2}\ 
A_p (q^2)\  \overline{U}_p\ \left( S^\mu q^2 - S\cdot q q^\mu\ \right) U_p
\ \ \ ,
\label{eq:anamat}
\end{eqnarray}
where $A_p (q^2)$ denotes the anapole form factor (the anapole moment is
defined to be $A_p (0)$) of the proton.
While such electromagnetic contributions vanish for on-shell photons, 
they give rise to local operators involving the proton 
and $e$ (or whatever charged probe is involved in the process).
In QCD, the anapole moment and form factor of the proton
have been computed in $\chi$PT~\cite{KS93,HHM89,SaSp01,MK00,anahigh}, 
and are dominated by the 
$h_{\pi NN}^{(1)}$ coupling 
(assuming that it is not anomalously small compared to 
estimates based on naive dimensional analysis).
Further, the anapole moments of a few nuclei have been studied theoretically 
(e.g. Ref.~\cite{HLM01}).
The proton anapole form factor at LO in $\chi$PT is 
found to be ~\cite{SaSp01,MK00}
\begin{eqnarray}
A_p (q^2) & = & 
+ {e g_A h_{\pi NN}^{(1)} M_N^2\over 48 \pi f}\  \tilde A_{\pi^+}(q^2)
\nonumber\\
\tilde A_{\pi^+}(q^2) & = & \tilde A (q^2, m_{\pi^+}) \ =\  
{12\over \left(\sqrt{-q^2}\right)^3}
\left[\ \left( m_{\pi^+}^2-{q^2\over 4}\right)
\tan^{-1}\left( {\sqrt{-q^2}\over 2 m_{\pi^+}}\right)
- {m_{\pi^+} \sqrt{-q^2}\over 2}\right]
\nonumber\\
& \rightarrow &  {1\over m_{\pi^+}} + {q^2\over 20 m_{\pi^+}^3}
\ +\ ... 
\ \ \ ,
\label{eq:anaQCD}
\end{eqnarray}
where the $q^2\rightarrow 0$ limit reproduces the anapole moment calculation
of Ref.~\cite{HHM89}.

\begin{figure}[!ht]
\centerline{{\epsfxsize=3.5in \epsfbox{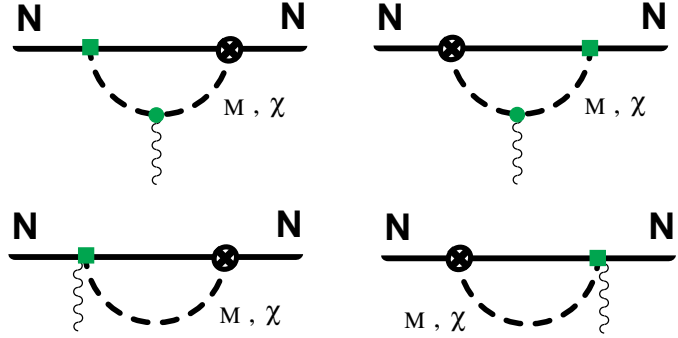}}} 
\vskip 0.15in
\noindent
\caption{\it 
One-loop graphs that give the leading contribution 
to the anapole moment and form factor of the nucleon.
A solid and dashed line denote a
{\bf 70}-nucleon and a meson, respectively.
The solid-squares denote an axial coupling given in eq.(\ref{eq:ints}),
while the crossed  circle denotes an insertion of the parity-violating
pion-nucleon  operators with coefficients $W_{1,2}$ in eq.~(\ref{eq:wkints}).
}
\label{fig:ana}
\vskip .2in
\end{figure}

While lattice computations of the proton anapole form factor are 
of somewhat secondary importance compared 
to the computation of $h_{\pi NN}^{(1)}$, it would be extremely
interesting to have a lattice QCD determination.
To this end, we present expressions for the anapole moment of the proton 
in two-flavor PQQCD.  
The LO contribution to the proton anapole form factor, arising from the
one-loop diagrams shown in Fig.~\ref{fig:ana}, is
\begin{eqnarray}
A_p^{(PQ)} (k^2) & & =\  
+ {e M_N^2\over 48 \pi f}\  
\left[\ g_A\ 
\left[\ 
{1\over 3}\left(2W_1-W_2\right) \ \tilde A_{ud}
\right.\right.\nonumber\\
& & \left.\left.\qquad
\ +\ W_1\ \left[ {1\over 6}+{1\over 4} q_j h_j - {1\over 4} q_j 
- {1\over 6} h_j
\right]
\left( \tilde A_{ju}-\tilde A_{uu}\ \right)
\right.\right.\nonumber\\
& & \left.\left.\qquad
\ +\ 
W_1\ \left[ {1\over 6}+{1\over 4} q_l h_l - {1\over 4} q_l - {1\over 6} h_l
\right]
\left( \tilde A_{lu}-\tilde A_{ud}\ \right)
\ \ \right]
\right.\nonumber\\
& & \left.
\ +\ g_1 \ \left[\ 
{1\over 72}\left( W_1\left( 4 \tilde A_{ju} 
+ \tilde A_{ld} + 4 \tilde A_{lu} - 5 \tilde A_{ud} - 
4 \tilde A_{uu} - \tilde A_{dd} + \tilde A_{jd}\right)
\right.\right.\right.\nonumber\\
& & \left.\left.\left.\qquad\qquad
\ +\ W_2 \left( 4 \tilde A_{ju} + 4 \tilde A_{ld} 
+ 4 \tilde A_{lu} - 8 \tilde A_{ud} - 
4 \tilde A_{uu} - 4 \tilde A_{dd} + 4 \tilde A_{jd}\right)\right)
\right.\right.\nonumber\\
& & \left.\left.
+\ {h_j\over 72}\left( W_1\left(4 \tilde A_{uu} 
- \tilde A_{ud} - 4 \tilde A_{ju} + \tilde A_{jd}\right)
+ 4 W_2\left(  \tilde A_{uu} -  \tilde A_{ud} 
-  \tilde A_{ju} + \tilde A_{jd}\right)\right)
\right.\right.\nonumber\\
& & \left.\left.
+\ {h_l\over 72}\left( W_1\left(4 \tilde A_{ud} 
- \tilde A_{dd} - 4 \tilde A_{lu} + \tilde A_{ld}\right)
+ 4 W_2\left(  \tilde A_{ud} -  \tilde A_{dd} 
-  \tilde A_{lu} + \tilde A_{ld}\right)\right)
\right.\right.\nonumber\\
& & \left.\left.
+\ {q_j\over 24}\left(  W_1\left( 2 \tilde A_{uu} - \tilde A_{ud} 
- 2  \tilde A_{ju} + \tilde A_{jd}\right)
+ W_2 \left( 2 \tilde A_{uu} - 4 \tilde A_{ud} 
- 2  \tilde A_{ju} + 4\tilde A_{jd}\right)\right)
\right.\right.\nonumber\\
& & \left.\left.
+\ {q_l\over 24}\left(  W_1\left( 2 \tilde A_{ud} - \tilde A_{dd} 
- 2  \tilde A_{lu} + \tilde A_{ld}\right)
+ W_2 \left( 2 \tilde A_{ud} - 4 \tilde A_{dd} 
- 2  \tilde A_{lu} + 4\tilde A_{ld}\right)\right)
\right.\right.\nonumber\\
& & \left.\left.
+\ {q_j h_j\over 24}\left(  W_1\left( 2 \tilde A_{ju} + \tilde A_{jd} 
- 2  \tilde A_{uu} - \tilde A_{ud}\right)
+ 2 W_2 \left( \tilde A_{ju} + 2 \tilde A_{jd} 
-  \tilde A_{uu} -2\tilde A_{ud}\right)\right)
\right.\right.\nonumber\\
& & \left.\left.
+\ {q_l h_l\over 24}\left(  W_1\left( 2 \tilde A_{lu} + \tilde A_{ld} 
- 2  \tilde A_{ud} - \tilde A_{dd}\right)
+ 2 W_2 \left( \tilde A_{lu} + 2 \tilde A_{ld} 
-  \tilde A_{ud} -2\tilde A_{dd}\right)\right)
\right]
\right.\nonumber\\ & & \left.\qquad
\ \right]
\ ,
\label{eq:anaPQ}
\end{eqnarray}
where we have used the short-hand $\tilde A_x = \tilde A_x (q^2)$ and we
have used the electric-charge matrix in PQQCD, 

\begin{eqnarray}
{\cal Q}^{(PQ)} & = & 
{\rm diag}\left(\ +{2\over 3}\ ,\  -{1\over 3}
\ ,\ q_j\ ,\  q_l\ ,\  q_j\ ,\  q_l\ \right)
\ \ \ ,
\label{eq:PQcharge}
\end{eqnarray}
where $q_j$ and $q_l$ are arbitrary charges.
The form factor in eq.~(\ref{eq:anaPQ}) clearly reduces to the QCD expression 
in eq.~(\ref{eq:anaQCD}) in the appropriate limit.

\section{Conclusions}
\label{sec:conc}

With a new generation of experiments designed to determine the PV
nucleon-nucleon interaction, it is time for a lattice QCD calculation
of the flavor-conserving PV coupling $h_{\pi NN}^{(1)}$.
Near future lattice simulations will be partially-quenched with quark masses
significantly larger than those of nature. 
Therefore we have presented the one-loop level
expressions required for the extraction of $h_{\pi NN}^{(1)}$, both with
vanishing pion four-momentum and with lattice kinematics.
Furthermore, we have --with extreme optimism-- given the one-loop expressions
necessary to determine the proton anapole moment and form factor.

\bigskip\bigskip

\acknowledgements

We thank Steve Sharpe for very illuminating discussions and critical comments 
on this work. 
This work is supported in
part by the U.S. Department of Energy under Grants No.  DE-FG03-97ER4014.

\end{document}